\documentclass[aps,preprint]{revtex4}%
\usepackage{amsfonts}
\usepackage{amsmath}
\usepackage{amssymb}
\usepackage{graphicx}%
\begin{document}
\title{Generation of spatially-separated spin entanglement in a triple quantum dot system}
\author{Ping Zhang, Qi-Kun Xue, }
\affiliation{International Center for Quantum Structures, Institute of Physics, Chinese
Academy of Sciences, Beijing 100080, China}
\author{Xian-Geng Zhao}
\affiliation{Institute of Applied Physics and Computational Mathematics, Beijing 100088, China}
\author{X.C. Xie}
\affiliation{Department of Physics, Oklahoma State University, Stillwater, OK 74075}
\pacs{PACS number}

\begin{abstract}
We propose a novel method for the creation of spatially-separated spin
entanglement by means of adiabatic passage of an external gate voltage in a
triple quantum dot system.%
%TCIMACRO{\TeXButton{TeX field}{\newline}}%
%BeginExpansion
\newline
%EndExpansion
%TCIMACRO{\TeXButton{TeX field}{\newline}}%
%BeginExpansion
\newline
%EndExpansion

\end{abstract}
\maketitle

Entanglement is of great interest in quantum computation\cite{Ekert,Nie},
quantum teleportation\cite{Ben}, and fundamental tests of quantum mechanics
\cite{Bell,Gree}. Designing and realizing quantum entanglement is extremely
challenging due to the intrinsic decoherence, which is caused by the
uncontrollable coupling with environmental degrees of freedom. A variety of
physical systems have been chosen to investigate the controlled, entangled
states. Among these are trapped ions\cite{Cirac}, spins in nuclear magnetic
resonance (NMR)\cite{Ger}, cavity-quantum-electrodynamics systems\cite{Dom},
Josephson junctions\cite{Mak}, and quantum dots (QDs)\cite{Loss}.

Due to the potential scalability and long decoherence times of the electron
spins, the solid-state quantum dot (QD) system has been extensively studied
for realization of a quantum computer both from theory and
experiment\cite{Aws}. How to extract entangled particles into
spatially-separated channels is an important issue. Costa and Bose\cite{Costa}
have shown how a QD can function as an effective beam splitter to entangle two
conduction electrons. Oliver \textit{et al}.\cite{Oliver} have considered a
single-level QD with one input and two output leads. Via a detailed
calculation of the transition amplitudes in the $T$ matrix formalism, they
have shown that a nonlocal spin-singlet state at the output leads may be
produced by the Coulomb interaction. Very recently, Saraga and
Loss\cite{Saraga} have proposed a novel spin entangler consisting of three
coupled quantum dots. They have showed that such a device can separate two
entangled electrons and extract them into two distinct channels.

In this paper we propose a means to generate spatially-separated
spin-entangled electron pair by using an adiabatic passage method. Adiabatic
passage is a powerful tool for manipulating a quantum system from an initial
state to a target state and has been extensively used recently to create
coherent superpositions of atomic states\cite{Weiz} and photon
states\cite{Parkins}. Until now very little efforts have been devoted to
implement adiabatic passage with artificial atom---i.e., QD system---although
there has been important progress in manipulating quantum states of the QD by
using an electric or magnetic field. Such examples are optically triggered
single-electron turnstile\cite{Zre} and manipulation of dot spin
dynamics\cite{Aws}. Here we illustrate how to implement quantum state transfer
in a QD system by use of adiabatic passage of one single parameter of gate
voltage. In contrast to real atom system, the QD system allows for a precise
control over the energy level structure. In some sense the entangler behaves
like a quantum tweezer\cite{Niu}, which picks up two electrons from the source
lead, entangles and adiabatically transfers them into the two output channels.
We notice that adiabatic control of the single-particle wavefunction was
recently suggested for current transport through a triple QD system\cite{Ren}.

As in Ref.[14], the entangler, as schematically shown in Fig. 1(a), is a
triple QD system in the Coulomb blockade regime. The central dot is connected
to an input lead, while the left and right dots are coupled to two output
leads. We consider the case that there are \textit{two} excess electrons in
the central dot. These two electrons are from the input lead and their ground
state is a spin-singlet state. Then the aim of the entangler is to extract the
singlet from the central dot, by transferring one electron to the left dot and
the other one to the right dot, and finally transport them
\textit{simultaneously} into the output ports. The final step can be easily
achieved via the single-particle tunneling mechanism when the two electrons
are present in the left and right dots. Therefore the most important step is
how to transfer the two electrons into the two side dots
simultaneously\textit{ }in a controllable way, which is our goal in this
paper. By adiabatic manipulation of the gate voltage $V_{g}$ applied on the
central dot, we show that the two spin-singlet electrons localized initially
in the central dot can be transferred into the two side dots simultaneously,
with each dot only being occupied by one electron [schematically shown in Fig.
1(b)]. The evolution is rather around \textit{one} avoided crossing in the
eigenenergy spectrum than along the ground state manifold which is
characterized by a series of Coulomb-induced avoided crossings. Typical
non-adiabatic tunneling mechanism, which will lead to a leakage out of the
target state, is greatly suppressed by adiabatic passage. Thus the
spatially-separated spin-entangled state builds up robustly. Our proposal is
very simple and easy to implement in experiments. The only variable parameter
is the single-particle energy level of the central dot, which can be
conveniently controlled by the gate voltage applied on it.

Accounting only for the occupation of the lowest single-particle state of each
dot, the system is described by a three-site Hubbard model%
\begin{align}
H(t) &  =\varepsilon_{s}(d_{L\sigma}^{+}d_{L\sigma}+d_{R\sigma}^{+}d_{R\sigma
})+\varepsilon_{c}d_{C\sigma}^{+}d_{C\sigma}-w\sum_{\alpha=L,R\sigma
}(d_{\alpha\sigma}^{+}d_{C\sigma}+\text{H.c.})\tag{1}\\
&  +U_{s}(n_{L\uparrow}n_{L\downarrow}+n_{R\uparrow}n_{R\downarrow}%
)+U_{c}n_{C\uparrow}n_{C\downarrow},\nonumber
\end{align}
where $d_{C\sigma}^{+}$ or $d_{L\sigma}^{+}$ ($d_{R\sigma}^{+}$) creates a
spin-$\sigma$ electron on the central or left (right) dot. $\varepsilon_{s}$
and $\varepsilon_{c}$ are the on-site orbital energies for the side and
central dots, respectively. $U_{s}$ is the Coulomb repulsion of the side dots,
while $U_{c}$ is the same for the central dot. $w$ denotes hopping amplitude
between the side dots and central dot. We will see below that $U_{s}$ has no
relevance to the problem since the states for which the two electrons occupy
the same left or right dot are unwanted states and can be eliminated from the
evolution of the system. The initial state---i.e., two electrons are localized
in the central dot---is a spin-singlet state, as has been verified in
experiments\cite{Tar}. Since there are no spin-dependent terms in (1), the
subsequent system evolution is confined to the spin-singlet states. In the
spin-singlet subspace, the two-particle basis is given by
\begin{equation}
|S^{1}\rangle=LL=d_{L\downarrow}^{+}d_{L\uparrow}^{+}|0\rangle,\tag{2a}%
\end{equation}%
\begin{equation}
|S^{2}\rangle=LC=\frac{1}{\sqrt{2}}(d_{C\downarrow}^{+}d_{L\uparrow}%
^{+}-d_{C\uparrow}^{+}d_{L\downarrow}^{+})|0\rangle),\tag{2b}%
\end{equation}%
\begin{equation}
|S^{3}\rangle=CC=d_{C\downarrow}^{+}d_{C\uparrow}^{+}|0\rangle,\tag{2c}%
\end{equation}%
\begin{equation}
|S^{4}\rangle=RC=\frac{1}{\sqrt{2}}(d_{R\downarrow}^{+}d_{C\uparrow}%
^{+}-d_{R\uparrow}^{+}d_{C\downarrow}^{+})|0\rangle,\tag{2d}%
\end{equation}%
\begin{equation}
|S^{5}\rangle=RR=d_{R\downarrow}^{+}d_{R\uparrow}^{+}|0\rangle,\tag{2e}%
\end{equation}%
\begin{equation}
|S^{6}\rangle=LR=\frac{1}{\sqrt{2}}(d_{L\downarrow}^{+}d_{R\uparrow}%
^{+}-d_{L\uparrow}^{+}d_{R\downarrow}^{+})|0\rangle,\tag{2f}%
\end{equation}
where $|0\rangle$ is the empty state with no excess electrons on the
entangler. We set the zero of the on-site energy as $\varepsilon_{s}=0$. Thus
in the following $\varepsilon_{c}$ is defined with respect to $\varepsilon
_{s}$. In the spin-singlet subspace, the Hamiltonian (1) is rewritten as a
$6\times6$ matrix
\begin{equation}
H=\left(
\begin{array}
[c]{llllll}%
U_{s} & \sqrt{2}w & 0 & 0 & 0 & 0\\
\sqrt{2}w & \varepsilon_{c} & \sqrt{2}w & 0 & 0 & w\\
0 & \sqrt{2}w & 2\varepsilon_{c}+U_{c} & \sqrt{2}w & 0 & 0\\
0 & 0 & \sqrt{2}w & \varepsilon_{c} & \sqrt{2}w & w\\
0 & 0 & 0 & \sqrt{2}w & U_{s} & 0\\
0 & w & 0 & w & 0 & 0
\end{array}
\right)  .\tag{3}%
\end{equation}
The gate voltage is applied on the central dot to control $\varepsilon_{c}$.
The starting state is $CC$ and the target state is spatially-separated state
$LR$. Given an initial state $|\Psi(0)\rangle$, the consequent time evolution
of the state is given by the Schr\"{o}dinger equation%
\begin{equation}
i\frac{d}{dt}|\Psi(t)\rangle=H|\Psi(t)\rangle.\tag{4}%
\end{equation}
The state $|\Psi(t)\rangle$ is expressed as a superposition of the six basis
states
\begin{equation}
|\Psi(t)\rangle=\sum_{k}c_{k}(t)|S^{k}\rangle.\tag{5}%
\end{equation}
The probability of finding target state $LR$ is given by $\rho_{LR}%
(t)=|c_{6}(t)|^{2}$.

In the absence of hopping term ($w=0$), the Hamiltonian (3) is diagonal and
the basis states are exact eigenstates. For illustration, the eigen-energies
are shown in Fig. 2(a) as a function of $\varepsilon_{c}$, which is tuned by
external gate voltage $V_{g}$. One can see that at
\begin{equation}
\varepsilon_{c}=-U_{c}/2, \tag{6}%
\end{equation}
the eigenstates $CC$ and $LR$ are degenerate. There are also the other three
level crossings, which have no relevance to this paper. The presence of the
hopping term $w$ opens up energy gaps at the crossings [see Fig. 2(b)], which
implies strong mixing between the corresponding hopping-free states.
Meanwhile, the location of the avoided crossing between $CC$ and $LR$ states
deviates from the expression in Eq. (3) [see the inset of Fig.2(b)]. To
illustrate the system dynamics at this avoided crossing, starting from the
initial state $CC$, we plot in Fig. 3(a) (solid lines) time evolution of the
probability of finding states $|S^{k}\rangle$ ($k=1,...,6$). One can see that
the dynamics is dominated by a resonant oscillation between $CC$ and $LR$,
while populations of other two-particle states are negligible. If the gate
voltage that controls $\varepsilon_{c}$ is suddenly switched to another value
at time that the occupation pprobability of state $LR$ is unity, then the two
electrons will cease oscillation and remain in the target state $LR$, as shown
in Fig. 3(b).

To find an analytical expression for the location of the avoided crossing in
the inset of Fig. 2(b), and the resonant oscillation period in Fig. 3, we
present a two-state approximation by adiabatically eliminating contributions
of the other states to the dynamics. From the expression of Hamiltonian (3)
one can see that the states $CC$ and $LR$ only couple with the states $LC$ and
$RC$ in which one electron occupies the central dot and the other on one side
dot. Because population of the states $LC$ and $RC$ remains very small during
time evolution as shown in Fig. 3, we can approximate $c_{2}(t)$ and
$c_{4}(t)$ in Eq. (5) to the first order of the hopping term:%
\begin{equation}
c_{2}(t)=-\frac{\sqrt{2}w}{\varepsilon_{c}}c_{3}(t)-\frac{w}{\varepsilon_{c}%
}c_{6}(t), \tag{7a}%
\end{equation}%
\begin{equation}
c_{4}(t)=-\frac{\sqrt{2}w}{\varepsilon_{c}}c_{3}(t)-\frac{w}{\varepsilon_{c}%
}c_{6}(t), \tag{7b}%
\end{equation}
while states $LL$ and $RR$ are completely neglected since their contributions
are very small (see Fig. 3). Substituting Eqs. (7) into the Schr\"{o}dinger
equation we reduce the system to an effective two-level system (TLS). The
reduced two-dimensional equation has the form%

\begin{equation}
i\frac{d}{dt}c_{3}(t)=(2\varepsilon_{c}+U_{c}-\frac{4w^{2}}{\varepsilon_{c}%
})c_{3}(t)-\frac{2\sqrt{2}w^{2}}{\varepsilon_{c}}c_{6}(t), \tag{8a}%
\end{equation}%
\begin{equation}
i\frac{d}{dt}c_{6}(t)=-\frac{2\sqrt{2}w^{2}}{\varepsilon_{c}}c_{3}%
(t)-\frac{2w^{2}}{\varepsilon_{c}}c_{6}(t). \tag{8b}%
\end{equation}
Therefore, starting from the initial state $CC$, the subsequent evolution is
featured by a coherent population transfer between the states $CC$ and $LR$,
just as shown in Fig. 3. In particular, when $\varepsilon_{c}$ is modulated to
satisfy the following equation:%
\begin{equation}
2\varepsilon_{c}+U_{c}-\frac{4w^{2}}{\varepsilon_{c}}=\frac{2w^{2}%
}{\varepsilon_{c}}\text{,} \tag{9}%
\end{equation}
the coherent transfer will be complete---i.e., population probability of
target state $|S^{6}\rangle$ can reach unity. The value of $\varepsilon_{c}$
that satisfies Eq. (9) corresponds to the exact location of the avoided
crossing shown in the inset of Fig. 2(b) and can be easily obtained:%
\begin{equation}
\varepsilon_{c}=\varepsilon_{c}^{0}=\frac{-U_{c}-\sqrt{U_{c}^{2}+16w^{2}}}{4}.
\tag{10}%
\end{equation}
Correspondingly, the resonance frequency of the two states is given from (8)
by%
\begin{equation}
\omega_{r}=-\frac{2\sqrt{2}w^{2}}{\varepsilon_{c}^{0}}\simeq\frac{4\sqrt
{2}w^{2}}{U_{c}}. \tag{11}%
\end{equation}
Thus the oscillation period of the population is given by
\begin{equation}
T=\pi/\omega_{r}\simeq\pi U_{c}/(4\sqrt{2}w^{2}). \tag{12}%
\end{equation}
The result of our two-state approximation is shown in Fig. 3 (dotted line).
Clearly, in comparison with the exact numerical solution, our two-state
approximation describes the system evolution very well. Therefore we arrive at
the conclusion that at time given by Eq. (12), a sudden switch of gate voltage
to another value will preserve the target state $LR$ and spatially-separated
spin entanglement comes into being. For a typical vertically-coupled
semiconductor QD, the amplitude of $U_{c}$ is about $5$meV while
$w\sim0.05U_{c}$. Therefore from Eq. (12), the generation time of
spatially-separated spin-singlet state is about $130$ps, much shorter than the
typical spin decoherence time of $100$ns\cite{Kik}.

The above analysis shows that by a manipulation of the single-particle energy
level of the central dot, there will be an avoided crossing in the energy
spectrum, which reduces the system to an effective two-level system consisting
of $CC$ and $LR$. However, the preparation and preservation of the target
state $LR$ shown in Fig. 3(b) depends on a precise tailored switching of the
resonant gate voltage. Any deviation may lead to significant errors. Moreover,
a sudden switching of the gate voltage introduces high frequencies and
therefore a population of higher-lying excited states. To overcome this
shortcoming, a more practicable strategy for the experimental implementation
of the above quantum state transfer process is the use of adiabatic passage:
$\varepsilon_{c}$ of the central dot is engineered to be near $\varepsilon
_{c}^{0}$ and satisfies $\varepsilon_{c}<\varepsilon_{c}^{0}$. The initial
state is engineered to be $CC$ state. Then the gate voltage applied on the
central dot will increase $\varepsilon_{c}$ towards $\varepsilon_{c}^{0}$. As
demanded by the quantum adiabatic theorem, if the gate voltage increases
infinitesimally slow the system will remain in its adiabatic state by
transferring the two electrons from the central dot to the two side dots,
forming spatially-separated spin-entangled state $LR$. The main leakage
mechanism from the target state is due to the well-known Landau-Zener (LZ)
tunneling. For the applied gate voltage scanning at a constant speed $\alpha$,
let us estimate the probability for LZ tunneling. Since we consider the
problem with $\varepsilon_{c}$ controlled around the avoided crossing of
states $CC$ and $LR$. Thus we assume that in the absence of the gate voltage,
the single-particle energy level of central dot is given by $\varepsilon
_{c}^{0}$. Then in the presence of gate voltage, the energy level is given by%
\begin{equation}
\varepsilon_{c}=\varepsilon_{c}^{0}+eV_{g}=\varepsilon_{c}^{0}+\alpha t.
\tag{13}%
\end{equation}
Substituting Eq. (13) into Eq. (8) we obtain
\begin{equation}
i\frac{d}{dt}c_{3}(t)=(\frac{2w^{2}}{\varepsilon_{c}^{0}}+2\alpha
t)c_{3}(t)-\frac{2\sqrt{2}w^{2}}{\varepsilon_{c}^{0}}c_{6}(t), \tag{14a}%
\end{equation}%
\begin{equation}
i\frac{d}{dt}c_{6}(t)=-\frac{2\sqrt{2}w^{2}}{\varepsilon_{c}^{0}}%
c_{3}(t)-(\frac{2w^{2}}{\varepsilon_{c}^{0}}+2\alpha t)c_{6}(t), \tag{14b}%
\end{equation}
where we have neglected time dependence of the terms proportional to
$w^{2}/\varepsilon_{c}^{0}$. Eq. (14) is a standard expression for two-state
LZ tunneling model, which gives the tunneling probability $r$
\begin{equation}
r=e^{-\frac{4\pi w^{4}}{\alpha(\varepsilon_{c}^{0})^{2}}}. \tag{15}%
\end{equation}
To suppress the LZ tunneling, one can either increase the gap of the avoided
crossing or decrease $\alpha$ such that the quantum state transition is in the
adiabatic regime.

To illustrate this adiabatic passage process, we show in Fig. 4 the exact
evolution of the probability of finding states $CC$ and $LR$ as a function of
gate voltage for different values of the scanning rate $\alpha$ (solid lines).
The initial state in these three panels is chosen to be the lower eigenstate
on the left side of the avoided crossing shown in the inset of Fig. 2(b),
which is dominated by the $CC$ state. The inset of Fig. 4(a) shows the results
with exact initial $CC$ state. In the adiabatic case where $\alpha$ is very
small, a complete $CC\rightarrow LR$ transfer occurs after the gate voltage
cross the avoided crossing shown in the inset of Fig. 2, implying that the LZ
tunneling is greatly suppressed. When the value of $\alpha$ becomes large, the
LZ tunneling takes place partially and the outcome is a superposition of the
states $CC$ and $LR$. This is shown in Figs. 4(b)-(c). For comparison, we also
plot in Fig. 4 the result of two-state approximation (14) (dotted lines).
Clearly, our two-state approximation describes the LZ tunneling very well. It
reveals in the inset of Fig. 4(a) that although the initial $CC$ state is not
the exact adiabatic state, the target state $LR$ can be also produced with
probability almost unity. This is the essential point of this paper.

We emphasize that the generation process of the spatially-separated spin
singlet is dependent upon choice of initial state. The initial state chosen in
this paper is $CC$. It is not the ground state of the Hamiltonian (3) when
system parameters are engineered to be around $\varepsilon_{c}^{0}$. In fact,
from the evolution of the \textit{exact} ground state shown in Fig. 5, one can
see that when $\varepsilon_{c}$ is around $\varepsilon_{c}^{0}$, the ground
state is a superposition of $LC$ and $RC$, while the population of states $CC$
and $LR$ is near zero. Therefore the adiabatic evolution shown in Fig. 4 is
not confined on the ground manifold. If one wants to transfer the initial $CC$
state to the target $LR$ state along the ground state evolution, as suggested
by Saraga and Loss\cite{Saraga}, then the quantum state will have to overcome
a series of avoided crossings (see Fig. 2). Correspondingly, the final target
$LR$ state can only be reached via $LC\rightarrow LR$ and $RC\rightarrow LR$
transitions. Thus it will need a very long time to transfer the electrons to
the $LR$ state. Our approach, however, does not involve the participation of
states $LC$ and $RC$ during system evolution. This is the essential
distinction between our approach and that proposed in Ref. [14] (see Figs. 6
which plots generation process of spatially-separated entangled current by two approaches).

Although the $CC$ state is far from the ground state structure when the value
of $\varepsilon_{c}$ is chosen to be near $\varepsilon_{c}^{0}$ and to satisfy
$\varepsilon_{c}<\varepsilon_{c}^{0}$, this does not mean that it cannot be
built up by the tunneling from the input lead into the central dot. In fact,
when the first electron tunnels into the central dot, due to the large energy
mismatch between $\varepsilon_{c}^{0}$ and $\varepsilon_{s}$, this excess
single electron cannot tunnel into the side dots. When the second electron
tunnels into the central dot, the off-resonance condition ($\varepsilon
_{c}<\varepsilon_{c}^{0}$) still ensures the prohibition of tunneling into the
side dots. Thus the two electrons can be robustly purified and form a
spin-singlet state $CC$, which is the first step of the entangler. Therefore
we would like to say that, unlike a many-body system which always favors the
ground state configuration, \textit{the few-body QD system can be conveniently
engineered to a localized state as a starting state due to its discrete energy
spectrum property which prohibits diffusion to ground state}. After the
initial $CC$ state builds up, the next step is to adiabatically apply a gate
voltage on the central dot, drive the system to target state $LR$ as shown in
the inset of Fig. 4(a). The final step is then to transport this spin-singlet
state to the two output ports, forming a spatially-separated spin-entangled
current. See Ref. [14] for details.

In summary, we have proposed a QD-based scheme for implementation of
spatially-separated spin entanglement. By use of a triple quantum dot it
becomes possible to prepare and preserve the spatially-separated spin-singlet
via an efficient adiabatic passage method. The evolution of the system is
restricted on an effective two-dimensional Hilbert space which consists of the
initial state $CC$ and the target state $LR$. Our approach, which is based on
a combination of eigen-energy spectrum analysis and adiabatic elimination of
dark states, may highlight the physical prospects in preparing entangled spin
qubits in QD\ systems.

This work has been supported by NSF and MOST of China, and by US-DOE.

{\Large Figure captions}

Fig. 1. (a) Model setup of a triple quantum dot entangler. There are two
excess spin-singlet electrons in the central dot. By engineering the central
dot level $\varepsilon_{c}$, the two electrons can simultaneously tunnel into
the two side dots with each electron occupying one dot. (b) The energy-level
schematic of the QD. When $\varepsilon_{c}$ is modulated to satisfy Eq. (12)
(see text), then the two electrons will be resonantly transferred into the two
side dots. The dashed arrows indicates the other spin configuration in the
central dot.

Fig. 2. The energy spectrum of a triple quantum dot as a function of the
single-particle energy level $\varepsilon_{c}$ for the values of $U_{c}%
=U_{s}=5$meV and (a) $w=0$, (b) $w=0.25$meV. The small rectangular part is
magnified in the inset to show the avoided crossing between states $CC$ and
$LR$.

Fig. 3. (a) Time evolution of the initial state $CC$ for the values of
parameters chosen to be at the crossing shown in the inset of Figure 2(b); (b)
Generation dynamics of $LR$ by switching $\varepsilon_{c}$ to another value at
time expressed by Eq. (12). Parameters are the same as in Fig. 2(b).

Fig. 4. Probability of finding $CC$ state and $LR$ state as the function of
the gate voltage $V_{g}$ for three values of scanning speed $\alpha$ of
$V_{g}$: (a) $\alpha=0.1$meV/ps; (b) $\alpha=1$meV/ps; (c) $\alpha=2$meV/ps.
The solid lines are the exact numerical results and dotted lines correspond to
the approximate results from Eq. (14). The initial state in three panels is
chosen to be the lower eigenstate on the left side of the avoided crossing
shown in the inset of Fig. 2(b), which is dominated by the $CC$ state. The
initial state in the inset of panel (a) is the exact $CC$ state.

Fig. 5. Probability of finding six two-particle basis states in the ground
ground state as a function $\varepsilon_{c}$. Dotted line gives location of
$\varepsilon_{c}^{0}$

Fig. 6. Quantum state evolution process in the entangler: (a) The ground state
evolution process proposed in Ref.[13]; (b) The adiabatic passage process
proposed in this paper. Here $C$ or $L$ ($R$) denotes one excess electron in
the central or left (right) dot.

\end{document}